\documentclass[12pt]{iopart}
\usepackage{graphicx}
\usepackage{iopams}
\usepackage[square,numbers,sort&compress]{natbib}
\usepackage{aas_macros}
\providecommand{\newblock}  

\newcommand{\ltot}{\ensuremath{L_{\rm M31,tot}}}
\newcommand{\lpt}{\ensuremath{L_{\rm M31,pt}}}

\begin{document}

\title[Unidentified high energy sources]{A luminosity constraint on the origin of unidentified high energy sources}

\author{J. M. Siegal-Gaskins$^{1,2}$, V. Pavlidou$^{1,3\footnote{Einstein (GLAST) Fellow}}$, A. V. Olinto$^1$, C. Brown$^1$ and B. D. Fields$^4$}
\address{$^1$ University of Chicago, Chicago, IL 60637}
\address{$^2$ Center for Cosmology and Astro-Particle Physics, The Ohio State University, Columbus, OH 43210}
\address{$^3$ Astronomy Department, California Institute of Technology, Pasadena, CA 91125}
\address{$^4$ University of Illinois at Urbana-Champaign, Urbana, IL 61801}
\ead{jsg@mps.ohio-state.edu}

\begin{abstract}
The identification of point sources poses a great challenge for the high energy community.  We present a new approach to evaluate the likelihood of a set of sources being a Galactic population based on the simple assumption that galaxies similar to the Milky Way host comparable populations of gamma-ray emitters.  We propose a luminosity constraint on Galactic source populations which complements existing approaches by constraining the abundance and spatial distribution of any objects of Galactic origin, rather than focusing on the properties of a specific candidate emitter.  We use M31 as a proxy for the Milky Way, and demonstrate this technique by applying it to the 
unidentified {EGRET} sources.  We find that it is highly improbable that the majority of the unidentified {EGRET} sources are members of a Galactic halo population (e.g., dark matter subhalos), but that current observations do not provide any constraints on all of these sources being Galactic objects if they reside entirely in the disk and bulge.  Applying this method to upcoming observations by the Fermi Gamma-ray Space Telescope has the potential to exclude association of an even larger number of unidentified sources with any Galactic source class.  
\end{abstract}

\section{Introduction}
\label{intro}

The Energetic Gamma Ray Experiment Telescope ({EGRET}) measured gamma-ray emission at energies greater than 100 MeV across the entire sky.  
More than half of the sources in the third {EGRET} catalog \citep{hartman_etal_99} were unidentified at the time of publication, and since then only a handful of those sources have been associated with known low-energy counterparts.  
Theoretical work has produced many candidate sources for these detections, including known Galactic source classes not previously confirmed as gamma-ray emitters such as microquasars \citep{paredes_etal_00}, and newly proposed populations of high energy Galactic sources such as dark matter annihilation in subhalos~\citep[e.g.,][]{berezinsky_etal_03,bergstrom_etal_99,blasi_olinto_tyler_03,calcaneoroldan_moore_00,tasitsiomi_olinto_02,taylor_silk_03,ullio_etal_02} and in mini-spikes around intermediate mass black holes \citep{bertone_etal_05}.  

The most direct approach to the confident association of gamma-ray sources with suitable counterparts is multiwavelength follow-up of individual objects
\citep[e.g.,][]{bloom_97, zook_97, gehrels_michelson_99,crb_99,tornik_02,fegan_05, tdr_05, lapalombara_etal_06, paredes_etal_08, reimer_funk_07}.
However, the sources in the third {EGRET} catalog typically have large error boxes, making association with known sources by positional coincidence difficult, and the large number of unidentified sources makes multiwavelength follow-up for every object impractical.
   
\begin{figure}[t]
\centering
 \includegraphics[width=0.6\textwidth]{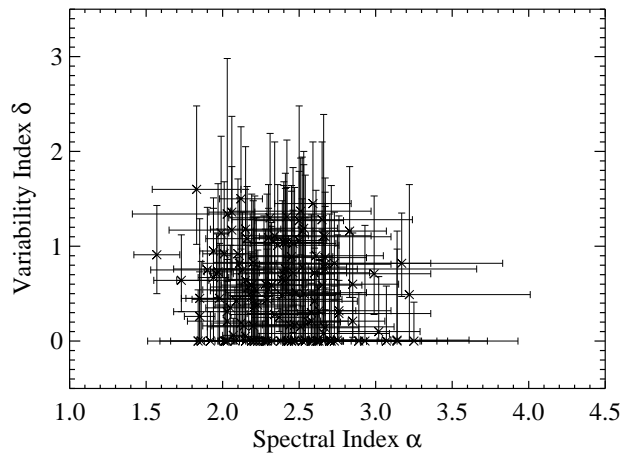}
\caption{Spectral and variability indices and associated uncertainties for sources originally unidentified in the 3EG catalog.  Spectral data is as published in the 3EG catalog; variability data is taken from \citet{nolan_etal_03}.\label{fig:variability}}
\end{figure}

Spectral and variability information can help by strengthening or excluding possible associations~\citep[e.g.,][]{merck_96,roberts_etal_05}.  However, large uncertainties in {EGRET's} spectral and variability data\footnote{Note that here we refer to general variability properties, such as variability indices, rather than correlated multiwavelength variability which, if present, can confirm the identification of a gamma-ray source with a lower-energy counterpart.}, particularly for the faintest sources, hinder the use of this information to extract sub-populations within the unidentified sources
\citep[e.g.,][]{grenier_00, reimerASSL01, grenier03, zhang_etal_04}. Figure~\ref{fig:variability} shows the distribution of the initially unidentified {EGRET} sources in the parameters $\delta$ (variability index from \citep{nolan_etal_03}) and $\alpha$ (spectral index from the 3EG catalog) with associated uncertainties.  
Although in principle association of sources with members of some source classes may be possible using correlations in these parameters, from Figure~\ref{fig:variability} it is clear that any clustering in the $\alpha-\delta$ plane would be significantly smeared by the measurement uncertainties. For this reason, using such clustering to pick out distinct classes of gamma-ray emitters is very difficult with the {EGRET} data, although future experiments that can probe smaller variability timescales and better determine source spectra may be able to use this information to extract source populations.

The situation is less favorable for associating sources with members of theoretically-motivated classes of gamma-ray emitters without previously established detections.  The uncertainties inherent in the emission features of unconfirmed source classes make association of individual sources tentative at best.  Association of unidentified sources with members of new source classes by positional correlation with low-energy counterparts has been proposed \citep{stu_der_95, colaf_02}, but these techniques can produce spurious results \citep{torres_reimer_05}.
Furthermore, approaches to associating sources with new populations which rely on matching unidentified gamma-ray sources with objects detected in other wavelengths cannot be used if the candidate counterparts are not necessarily detectable in other energy ranges (e.g., the case of annihilating dark matter clumps).
While emission from dark matter annihilation may have unique signatures such as a ``smoking gun'' spectral line, for many scenarios it is unlikely that even next-generation experiments could detect such a signal.  

One way to make progress in making source associations that does not require any knowledge of candidate counterparts is the $\log N-\log S$ diagram, the distribution of the number of sources with observed flux \citep[e.g.,][]{reimer_thompson_01, reimerASSL01, grenier_01, grenier03, bhatta_03, reimer_05}.  However, variability, large uncertainties in measured fluxes, and the presence of multiple populations within the unidentified sources \citep{grenier_01} pose problems for this method.
Still, one advantage of using $\log N-\log S$ distributions is that they enable comparison between Galactic and extragalactic populations without requiring a large number of detailed assumptions.  While this is a population-level approach, rather than one which focuses on individual objects, it can help distinguish between Galactic and extragalactic populations, and thus provides information not only about the source populations themselves but also about possible contributions from unresolved members of the same source class to diffuse backgrounds \citep{pavlidou_siegal-gaskins_fields_etal_08}.

The source distance is perhaps the most obvious difference between extragalactic and Galactic populations, and is an essential ingredient in connecting the source luminosity, which is often motivated by theoretical arguments but not directly observable, with the measured flux.  Previous work has considered the expected distances and luminosities implied by associating unidentified sources with specific Galactic populations \citep{mukh_95,grenier03}.  However, a challenge for this approach is finding an appropriate ``standard candle'' with which to compare the derived luminosities.
In this work we revisit this type of approach, and introduce a novel standard candle with which we can compare the total luminosity of a proposed Galactic population: the total luminosity of M31\@. 
By assuming that the nearby galaxy M31 hosts gamma-ray emitting populations similar to those of the Milky Way, we determine whether the bulk of the unidentified sources in a given catalog can be of Galactic origin and place constraints on the spatial distribution of any proposed population without assuming a specific candidate emitter.  Our approach uses only angular position and flux information to evaluate the plausibility of Galactic distributions of these sources.  
In anticipation of the large source catalogs expected from the recently launched Fermi Gamma-ray Space Telescope\footnote{http://fermi.gsfc.nasa.gov} and other experiments in the near future, we demonstrate the potential of this technique by applying it to the existing {EGRET} data set.  

Confident association of high-energy sources with counterparts is a difficult problem involving many degeneracies.  As a result, individual constraints on the nature of detected sources are often weak.
For this reason, as many {\it independent} constraints as possible are needed to robustly determine the nature of the unidentified sources. The approach presented here, which uses M31 as a standard candle for the gamma-ray luminosity of normal galaxies, involves its own uncertainties which we discuss in detail below.  But, these uncertainties are independent of those of other methods, so in this way our proposed constraint adds a new piece to the puzzle. 
In \S\ref{constraints} we discuss the set of unidentified sources from the third {EGRET} catalog, outline our approach for placing constraints on unidentified source populations, and present our results.  We discuss the potential of this approach in the {\it Fermi} era in \S\ref{fermi} and conclude in \S\ref{conc}.

\section{Constraining a Galactic population with M31}
\label{constraints}

Like the Milky Way, M31 is a luminous, high surface brightness galaxy with a mass of $\sim10^{12}$ M$_{\odot}$.  Having formed and evolved in a similar environment,
M31 is also akin to the Milky Way in its structure and dynamical properties \citep[see, e.g.,][]{kzs_02}.
Because of this close resemblance, it seems likely that any gamma-ray emitting source populations (including both known astrophysical source populations and proposed populations such as dark matter subhalos) found in the Milky Way are also present in M31.  Under this assumption we test candidate Galactic populations by asking whether the total luminosity of a comparable population in M31 is consistent with current observational constraints on the total gamma-ray luminosity of M31.  For simplicity, we assume that all point sources in M31 are at precisely the distance of M31, i.e., the collective luminosity of the point sources can be compared to the total luminosity of M31.

{EGRET} did not detect M31, determining a 2$\sigma$ upper limit on the flux at the coordinates of M31 of $F_{(\ell,b)_{\rm M31}}<4.9\times 10^{-8}$ cm$^{-2}$ s$^{-1}$ \citep{hartman_etal_99}.  \citet{blom_etal_99} derive a more restrictive upper limit for M31 from archival {EGRET} data using a maximum likelihood technique and simultaneously fitting point source models of the M31 emission and nearby catalog sources.  Their analysis finds a 2$\sigma$ upper limit for the emission from M31 of $F_{\rm M31}(>$100 MeV$)<1.6\times 10^{-8}$ cm$^{-2}$ s$^{-1}$.  Since this upper limit is roughly 3 times smaller than the value given in \citet{hartman_etal_99}, we conservatively adopt the limit from \citet{hartman_etal_99}. 
We take the distance to M31 to be $d_{\rm M31}=785$ kpc \citep{mcconnachie_etal_05}, which gives the upper limit on the photon luminosity of M31 for energies greater than 100 MeV, 
\begin{equation}
\label{eqn:lhartman}
\ltot < 3.6\times10^{42} \ {\rm s}^{-1}\quad  \mbox{\it M31 total emission limit}  
\end{equation} 
or, in more convenient units,  $\ltot < 0.38$ (kpc/cm)$^{2}$ s$^{-1}$.  

In determining the appropriate limit to place on the luminosity of candidate source populations, we note that the entire gamma-ray luminosity of M31 cannot be attributed to point sources.  \citet{pavlidou_fields_01} calculate the expected flux of diffuse gamma-ray emission due to cosmic ray interactions to be $F_{\rm diff}(>$100 MeV$)\sim1.0\times10^{-8}$ cm$^{-2}$ s$^{-1}$, more than 20\% of $\ltot$.  Consequently, the expected luminosity of M31 due to point sources is restricted to 
\begin{equation}
\label{eqn:lhartmanpt}
\lpt \lesssim 2.9\times10^{42} \ {\rm s}^{-1}\quad \mbox{\it M31 point source limit}
\end{equation}
or $\lpt \lesssim$ 0.30 (kpc/cm)$^{2}$ s$^{-1}$.  
These upper limits constrain the gamma-ray emission of a point source within a solid angle set by EGRET's angular resolution.  For E $>$ 100 MeV, the fraction of photons contained within a cone of half-angle $\Theta_{\rm PSF} = 3.3^{\circ}$ is $f_{\gamma} = 0.69$, so the luminosity of a proposed population relevant for comparison with these upper limits is reduced by a factor of $f_{\gamma}$.
We use these luminosity limits to constrain the Galactic distribution of the unidentified sources by requiring that the total luminosity of the unidentified sources $L_{\rm tot}$ be consistent with these bounds, i.e.,  
\begin{equation}
f_{\gamma}L_{\rm tot}< L_{\rm M31}
\label{eqn:condition}
\end{equation}
taking $L_{\rm M31}$ to be one of the upper limits defined in Equations~\ref{eqn:lhartman} and \ref{eqn:lhartmanpt}.  We note that $\lpt$ is dependent on the adopted model for the truly diffuse emission, but we include it for completeness.  
 
Emission from unresolved point sources is another component of the Galactic luminosity which is measured as diffuse emission.  The contribution from unresolved sources may be relevant because our assumption that M31 and the Milky Way host similar populations of gamma-ray emitting sources requires that unresolved sources enhancing the Galactic diffuse emission also be present in M31.  The prediction we use for the diffuse flux of M31 only represents the expected genuinely diffuse emission, and does not account for the luminosity enhancement by unresolved point sources.  In particular, it is likely that unresolved members of the same class or classes as the unidentified sources are present and contribute to the diffuse luminosity of the Milky Way, if these populations are indeed Galactic.  However, in order to appropriately account for this emission, it would be necessary to make numerous assumptions about the properties of the proposed Galactic population, including the cumulative flux distribution extending below the flux sensitivity of {EGRET} and the location of the unresolved sources.  Without assuming a specific candidate emitter, we have no theoretical basis on which to construct either of these inputs.  Although one might attempt to determine the angular distribution of the unresolved sources based on observations of the Galactic diffuse flux, the distances (and hence the collective luminosity) of the unresolved sources would be highly model-dependent and sensitive to assumptions about the spatial distribution of the genuinely diffuse emission.  Consequently, due to large uncertainties in the magnitude of this emission, we again make the very conservative choice to ignore this guaranteed component when placing constraints, allowing all of the expected point source luminosity to be attributed to the resolved unidentified sources.

Additionally, there is already a guaranteed contribution to the luminosity of point sources from confirmed detections of members of known Galactic source classes, such as pulsars and supernova remnants, which further reduces the luminosity available to the unidentified sources.   
However, the total gamma-ray luminosity of these objects is very small 
compared to the diffuse luminosity of either the Milky Way or M31.
Including this component would not alter our results in any substantial way, so for
simplicity we ignore it.
 
In this work we ask whether the unidentified
sources (or a subset of them) could likely be members of a Galactic population.  
Since the flux, rather than the luminosity, of the unidentified sources is known, it is necessary to assume a distance to each source to determine the luminosity of a proposed population consisting of these sources.  Most candidate Galactic populations are likely to be associated with either the disk and bulge (e.g., pulsars, supernova remnants) or the dark matter halo (e.g., dark matter clumps, intermediate mass black holes), so in these cases we have a natural expectation for the spatial distribution of the sources.  In particular, for the case of the disk and bulge populations, we expect that the typical distances of the sources will reflect the spatial extent of the disk, roughly 30 kpc in radius (as in the emission model of \citep{hunter_bertsch_catelli_etal_97}).  For populations associated with the dark matter halo, we anticipate a distribution of the source distances out to at least the virial radius of the Galaxy $r_{\rm vir}$, which is 258 kpc in the favored model of \citet{kzs_02}, and note that recent numerical simulations find that the dark matter subhalo distribution extends significantly beyond this distance \citep{prada_klypin_simonneau_etal_06, diemand_kuhlen_madau_etal_08, springel_white_frenk_etal_08}.  

The angular distribution of sources can be an important discriminator between various Galactic distributions of sources and extragalactic sources.  Due to the relatively small offset of our position from the Galactic Center compared to the typical distances of sources associated with the dark matter halo, the angular distribution of such sources would appear largely isotropic.  In contrast, sources associated with the disk and bulge would be concentrated at low latitude.  However, due to EGRET's unequal exposure map, using isotropy information to constrain the EGRET unidentified sources requires careful study and is beyond the scope of this work.

\subsection{The unidentified sources}
\label{sources}

The third {EGRET} catalog contains 271 sources (E$>$100 MeV), 101 of which were
initially identified or associated with likely lower-energy counterparts by the {EGRET} team.  The majority of these sources are blazars, along with a smaller number of pulsars, the Large Magellanic
Cloud, a radio galaxy, and a solar flare.  An additional 48 of the originally unidentified sources subsequently have been suggested to be associated with plausible counterparts, including binaries, supernova remnants, gas clouds, microquasars, and black holes.  We evaluate the total luminosity of possible Galactic distributions of the unidentified sources using two subsets of sources from the {EGRET} catalog as our candidate populations, and emphasize that the hypothesis we are attempting to reject with the luminosity constraint is that all of the unidentified sources are of Galactic origin.

We first consider the set of all sources published without identifications in the catalog.  To avoid biasing our sample toward higher fluxes, for each source we use the P1234 flux (summed flux over cycles 1, 2, 3, and 4).  As a result, we exclude two additional sources which were not published with P1234 fluxes, leaving a total of 168 sources in this set, which we term the ``original sample''.

We also construct a more restricted source sample by excluding those sources for which plausible associations with low energy counterparts have been suggested since the catalog was published.  We make use of an updated listing of {EGRET} unidentified sources, and exclude from our sample all sources for which an association has been suggested, regardless of the significance of the association.  Associations included in this compilation are simply the results reported in recent publications; the validity of these suggested counterparts has not been evaluated by a single standard.  We also remove from this sample the few sources noted by the {EGRET} team as possible or likely artifacts.  After applying these restrictions, our ``restricted sample'' consists of 119 unidentified sources.  
Omitting these sources from our restricted sample is again a conservative choice because increasing the number of sources in our candidate population would lead to an increase in the total population luminosity, and hence any distribution would be more likely to violate the M31 luminosity constraint.  We comment that 
\citet{sowards-emmerd_romani_michelson_03,sowards-emmerd_romani_michelson_etal_04} report plausible associations of many high-latitude unidentified EGRET sources with blazars, and these associations are not reflected in our restricted sample.  We also note that a revised EGRET source catalog has been published by \citet{casandjian_grenier_08} that does not confirm the detections of several 3EG sources.  Although {\it Fermi} has now released a bright gamma-ray source list \cite{abdo_09}, we do not consider those sources in this study since they are high-confidence, high-flux sources, and most of the {\it Fermi} unidentified sources will likely be faint, resulting in larger positional error circles.

\subsection{Simple luminosity tests}
\label{simple} 

The {EGRET} data provides a measurement of the flux and the angular position of each source, so the distance to each unidentified source is needed to calculate the total luminosity of a candidate population: 
\begin{equation}
\label{eqn:ltot}
L_{\rm tot}=\sum_{i}L_{i}=4\pi\sum_{i}d_{i}^{2}F_{i}
\end{equation}
where the summation is over each unidentified source $i$ and the flux, luminosity, and distance of each source are denoted by $F_{i}$, $L_{i}$, and $d_{i}$ respectively.  
We begin by using some simple models to assign source distances.  These models are not realistic, but instead indicate the typical allowed distances of these objects from our position in the most extreme scenarios.    

We first consider the case that all unidentified sources are at a fixed distance $d_{0}$ from us, i.e., $d_{i}=d_{0}$, and use Equation~\ref{eqn:ltot} to determine the total population luminosity $L_{\rm tot}$ in terms of $d_{0}$.  We then ask what is the maximum distance $d_{0}$ at which we can place the sources without violating the M31 luminosity limit \ltot?  Because sources observed to have a given flux will be more luminous if located at a greater distance from the observer, the luminosity constraint $f_{\gamma} L_{\rm tot} \le \ltot$ limits the maximum distance:
\begin{equation}
\label{eqn:unidist}
d_{0}\le\sqrt{\frac{\ltot}{4\pi f_{\gamma} \sum_{i}F_{i}}}.
\end{equation}
Using the original source sample we find that the maximum distance $d_{0}=30$ kpc which is plausible for a disk and bulge source population, however, a population of halo sources would extend at least as far as $r_{\rm vir}=258$ kpc, and consequently would
have typical distances from our position much greater than 30 kpc.  This simple test implies that the unidentified sources are not primarily associated with the dark matter halo of the Galaxy.

We next consider a population of sources which all have the same intrinsic luminosity $L_{0}$.  We consider the original sample as above to set the individual source luminosity $L_{0}=L_{\rm tot}/N$, with $N=168$ sources, and set the total relevant population luminosity $f_{\gamma}L_{\rm tot}$ equal to the total luminosity upper limit $\ltot$.  Using the measured flux of each source, we calculate the corresponding distance to each source in this scenario,
\begin{equation}
\label{eqn:stdcandle}
d_{i}=\sqrt{\frac{L_{0}}{4 \pi F_{i}}}=\sqrt{\frac{\ltot}{4 \pi f_{\gamma} N F_{i}}}.  
\end{equation}
Under this assumption the sources are at a mean distance of 37 kpc and we find, as for the fixed distance case presented above, that the typical distances of the sources are uncomfortably small for a plausible halo population, but quite reasonable for a disk and bulge population.

Naturally, we now ask what is the \emph{maximum} average distance  $\bar{d} = \sum_{i}d_{i}/N$ allowed while still respecting the luminosity constraint?  We extremize $\bar{d}$ subject to the condition that our set of fluxes $F_{i}$ and assigned distances $d_{i}$ produce a given total population luminosity $L_{\rm tot}$ via Equation~\ref{eqn:ltot} (which we will take to be $\ltot/f_{\gamma}$), and find that the maximum average distance is given by
\begin{equation}
\label{eqn:dmax}
\bar{d}_{\rm max}=\frac{1}{N}\sqrt{\frac{\ltot}{4\pi f_{\gamma}}}\sqrt{\sum_{i}\frac{1}{F_{i}}}.
\end{equation}
For the original source sample and the total luminosity upper limit $\ltot$, $\bar{d}_{\rm max}=40$ kpc.  Once again, this distance is too small to be characteristic of a halo population, but consistent with a disk and bulge distribution.

\subsection{Assigning source distances}
\label{distances}

We now take a more sophisticated approach to construct our population's spatial distribution.  
Our method for assigning distances is motivated by the goal of testing proposed source populations, so here we determine the distance to each source by considering the expected Galactic distribution of various types of populations.  We emphasize, however, that alternative methods of assigning source distances can also be used to determine the luminosity of the candidate population, so the luminosity test we propose can be used in conjunction with existing approaches to provide an additional, complementary constraint.
In general, candidate Galactic populations can be classified as living either in the halo or in the disk and bulge.  Because these populations can be associated with a measured mass component of the Galaxy, we use mass density as a proxy for source density.  

For source populations expected to reside in the disk and bulge, we approximate the mass distribution with a Miyamoto-Nagai disk \citep{miyamoto_nagai_75} and a Hernquist bulge \citep{hernquist_90}.  We take the disk mass to be M$_{\rm d} = 4 \times 10^{10} $ M$_{\odot}$ \citep{kzs_02}, and the scale parameters $a$ and $b$ to be 6.5 kpc and 0.26 kpc respectively \citep{johnston_etal_96}.  For the bulge mass we use M$_{\rm b} = 8.0 \times 10^{9} $ M$_{\odot}$ ($m_{1} + m_{2}$ in \citep{kzs_02}), with the scale parameter $a=0.7$ kpc \citep{johnston_etal_96}.  The disk and bulge profiles are truncated at a radius of 30 kpc from the Galactic Center.  
 
For the case of candidate populations correlated with the dark matter distribution of the Galaxy, we model the dark matter halo of the Milky Way using a NFW profile \citep{nfw_95} with mass M$_{\rm halo} = 10^{12}$ M$_{\odot}$ and concentration $c = 12$ \citep{kzs_02}, truncating the profile at $r_{\rm vir}=258$ kpc from the Galactic Center.  We note that since $r_{vir}$ represents a sizable fraction of the distance between the Milky Way and M31,
the difference between the flux measured at the Milky Way from a source $r_{\rm vir}$ closer or farther from us than the distance of M31 is not negligible.  In other words, the extent of M31's source population along the line-of-sight could be large enough that our assumption that all sources associated with M31 are at precisely the distance of the center of M31 is a crude approximation.  The largest variations possible due to this effect are, however, small compared to the uncertainties inherent in our approach: in the most extreme cases an individual source's flux may be enhanced by up to a factor of $\sim 2.2$ or decreased by a factor of $\sim 1.8$.  For an isotropically distributed population, the overall change in the M31 luminosity is not significant.

For M31, the angle $\Theta_{\rm PSF}$ of $3.3^{\circ}$ corresponds to a projected radius of $\sim 45$ kpc, which encloses the expected emission region of the disk and bulge, but does not fully enclose the dark matter halo.  For the halo parameters used here, a cylinder with an angular radius of $3.3^{\circ}$ centered on M31 encloses a fraction $f_{\rm halo} \sim 0.38$ of the halo mass, so for a halo distribution we assume the total source luminosity within the cone reflecting EGRET's angular resolution will be further reduced by a factor $f_{\rm halo}$ beyond the factor $f_{\gamma}$ reflecting the photon containment fraction.

We use the P1234 fluxes and the angular positions of the sources as measured by {EGRET}.
Each source in the catalog is reported with an estimate of positional uncertainty, $\Theta_{95}$, typically $\sim$1$^\circ$.  
This value is for most objects the angular radius of a circle containing the same solid angle as the 95\% confidence level contour.  
However, in general the contour is not a circle but rather a complex shape, and so for simplicity we will consider $\Theta_{95}$ to be the angular
uncertainty of both the Galactic longitude $\ell$ and latitude $b$ measurements.  We note that \citet{mattox_etal_01} have provided elliptical contour fits to the {EGRET} sources, but  
because we do not expect the mass density to vary significantly over such small scales, 
generalizing the error contours to squares is a suitable approximation for our purposes.

\begin{figure}[t]
\centering
 \includegraphics[width=0.45\textwidth]{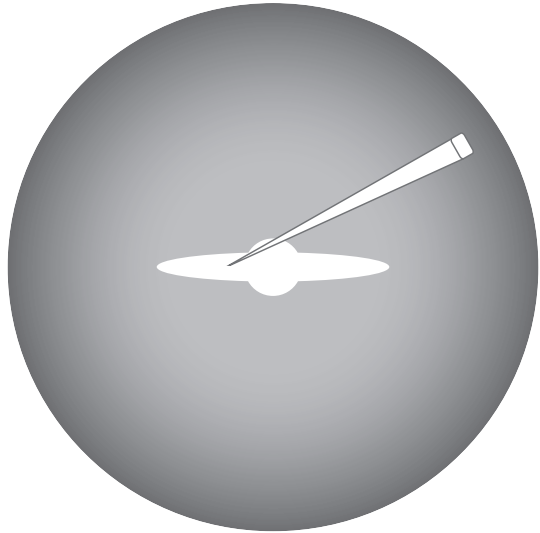}
\caption{We calculate the cumulative distribution function for the distance to each object by integrating the mass enclosed within the error boxes from the observer along the line of sight.\label{fig:los}}
\end{figure}

For each angular source position ($\ell,b$) we construct a cumulative distribution function for a given Galactic mass distribution to describe the likelihood of the source being within a particular distance along that line of sight.  The probability of a source with a given angular position being located within a distance $d$ from us is then given by
$\mathcal{P}(d)=M_{\rm encl}(d)/M_{\rm encl}(d_{\rm max})$,
where $d_{\rm max}$ is the distance along that line of sight at which we truncate our mass distribution to produce a finite volume in which to place the sources.
The mass enclosed along the line of sight within a solid angle    
defined by the positional error boxes and extending from the observer
out to a distance $d$ is given by 
\begin{equation}
M_{\rm encl}(d) = \int_{0}^{d} \int_{\ell_{-}}^{\ell_{+}}  \int_{b_{-}}^{b_{+}} \rho(z,\ell,b) z^{2} \cos(b) {\rm d}b {\rm d}\ell {\rm d}z, 
\end{equation}
where $\ell_{\pm}=\ell \pm \Theta_{95}$, $b_{\pm}=b \pm \Theta_{95}$, and $\rho$ is the mass density.  The variable $z$ describes integration along the line of sight.  This is illustrated schematically in Figure~\ref{fig:los}.

We use a Monte Carlo algorithm to generate realizations of the population distributions.  For each realization
a distance is assigned to each source by sampling the appropriate cumulative distribution function associated with the angular position of that source.  The total luminosity for the realization is then calculated.

\subsection{Results}
\label{results}

\begin{figure}[t]
\centering
\includegraphics[width=0.45\textwidth]{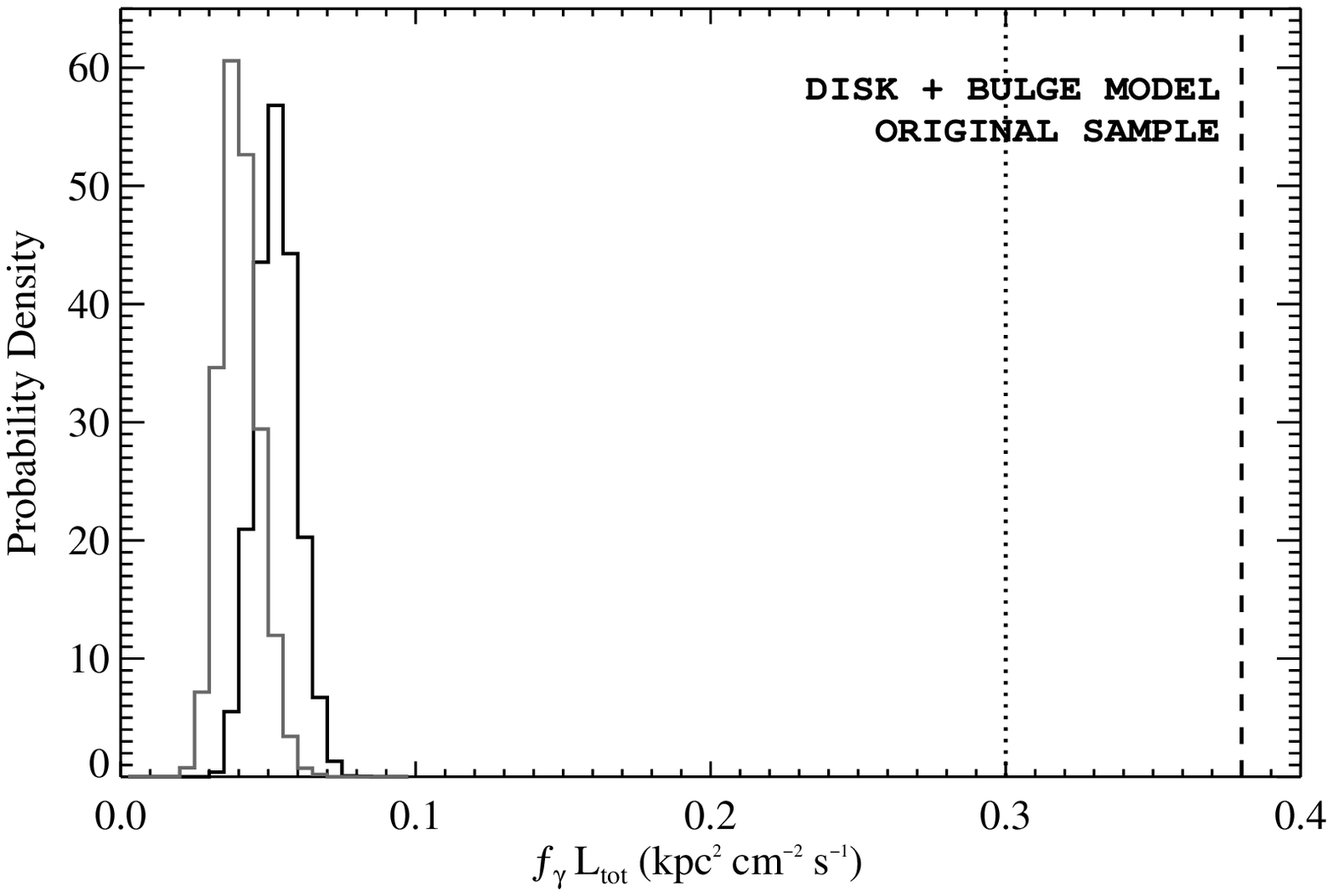}
\includegraphics[width=0.45\textwidth]{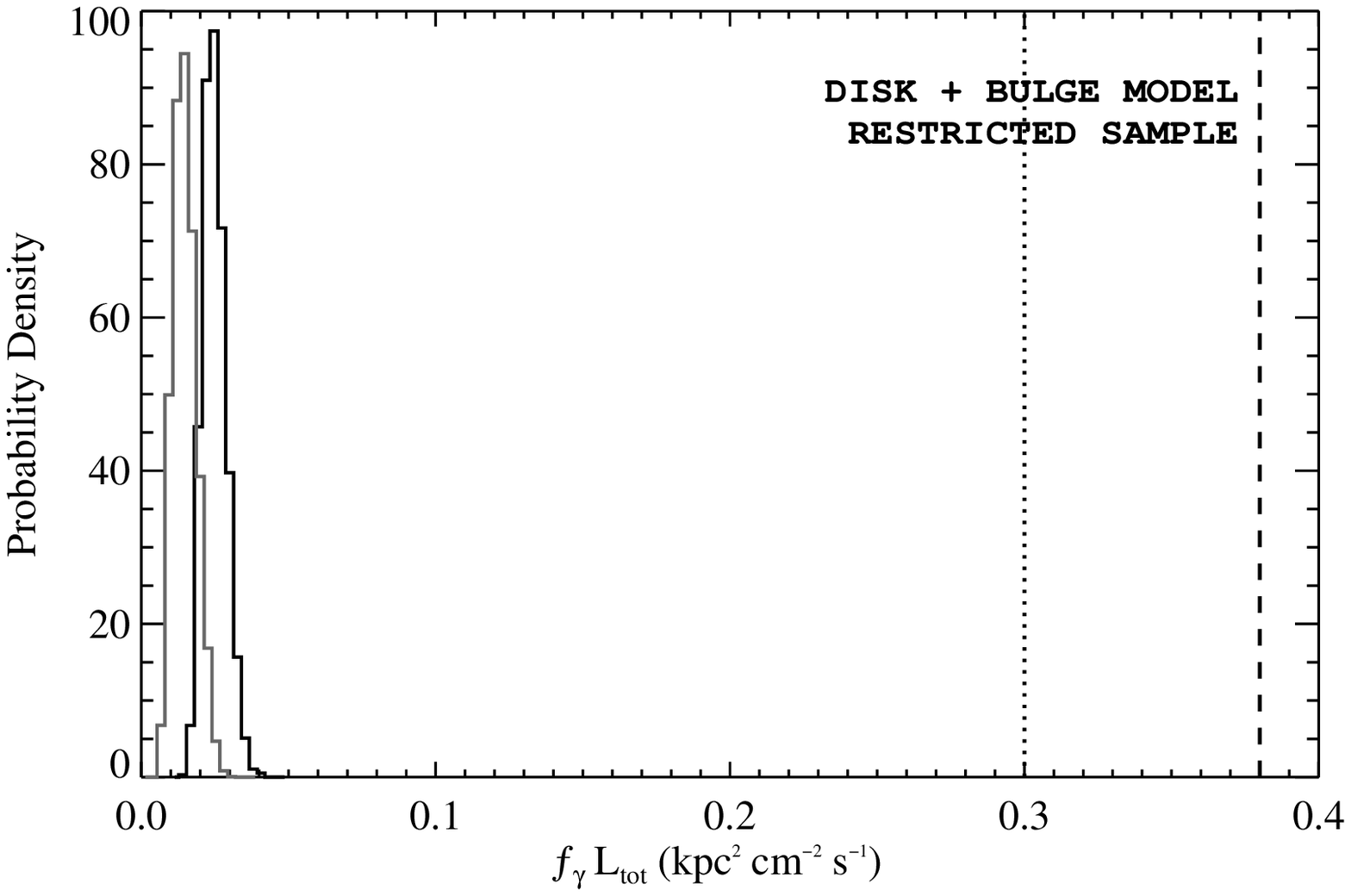}
\caption{Probability density functions for total luminosity of a population of sources associated with the disk and bulge.  The distribution of total luminosities for 5000 realizations, using sources with $|b|\le5^\circ$ ({\it light gray}) and using all sources ({\it black}), is shown for the original sample set ({\it left panel}) and the restricted sample set ({\it right panel}).   The M31 total luminosity upper limit $\ltot$ ({\it dashed}) and derived upper limit for the luminosity of point sources $\lpt$ ({\it dotted}) are shown for reference.\label{fig:dbresults}}
\end{figure}

Figure~\ref{fig:dbresults} shows the results of the luminosity test for disk and bulge distributions.  For each of our samples, we first construct our realizations using only the sources which roughly appear to be located in the Galactic plane ($|b|\le5^\circ$), and then using all sources.  Fitting these distributions to a Gaussian we find the mean of the distribution $\mu=0.040$ for the Galactic plane sources of our original sample, and $\mu=0.053$ for all sources in that sample, in units of (kpc/cm)$^{2}$ s$^{-1}$.  For the restricted sample, the mean values for the Galactic plane sources and for all sources are $\mu=0.014$ and $\mu=0.024$ respectively.  In all four cases $\ltot$ and $\lpt$ lie far above the total luminosity distributions, thus our luminosity test is consistent with the possibility that all of the unidentified sources are part of a Galactic population distributed in the disk and bulge.  We note that this finding is consistent with that of \citet{grenier_97}, in which a complementary approach based on isotropy arguments was taken to determine whether the unidentified sources could be associated with various Galactic structures.

\begin{figure}[t]
\centering
\includegraphics[width=0.45\textwidth]{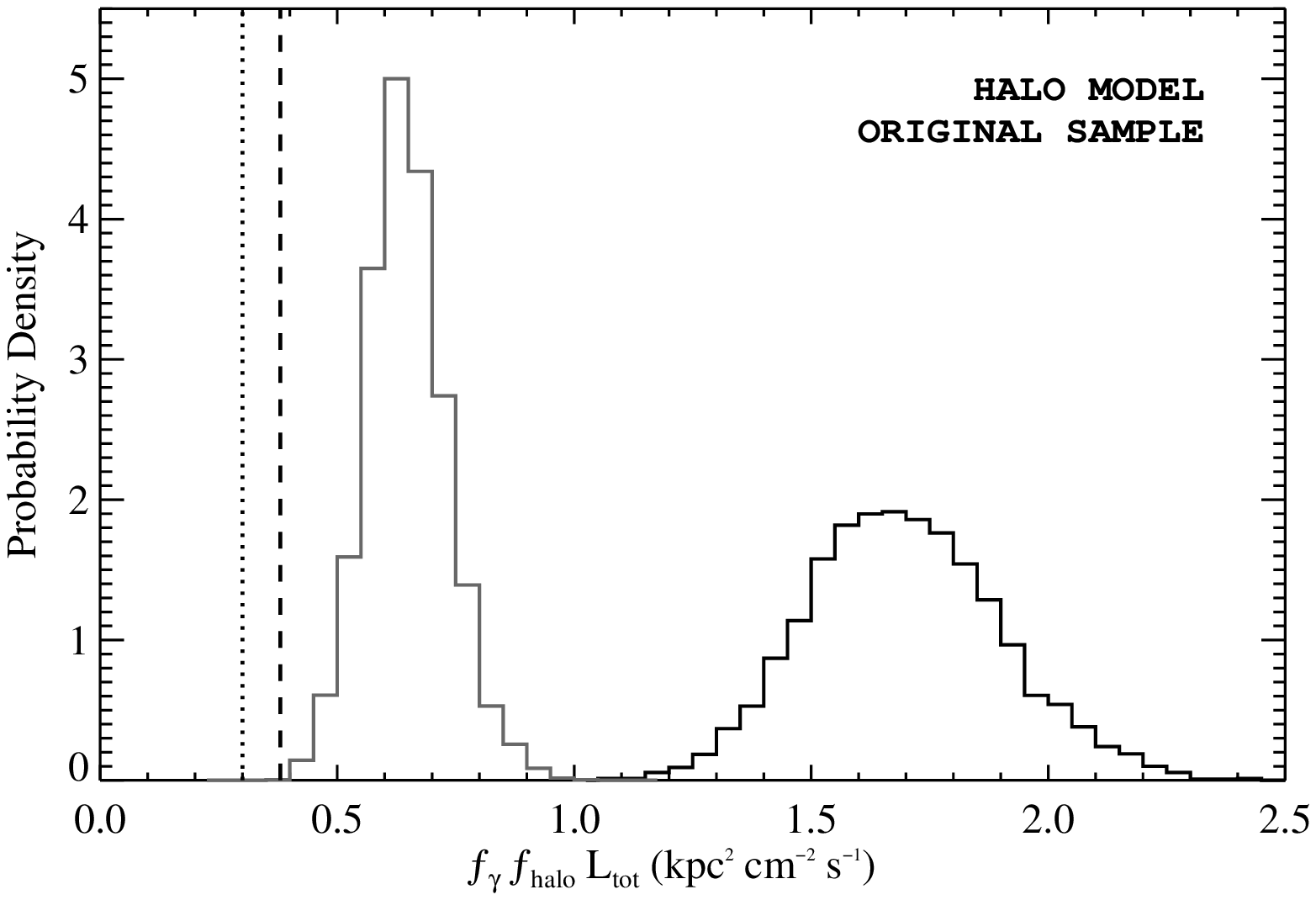}
\includegraphics[width=0.45\textwidth]{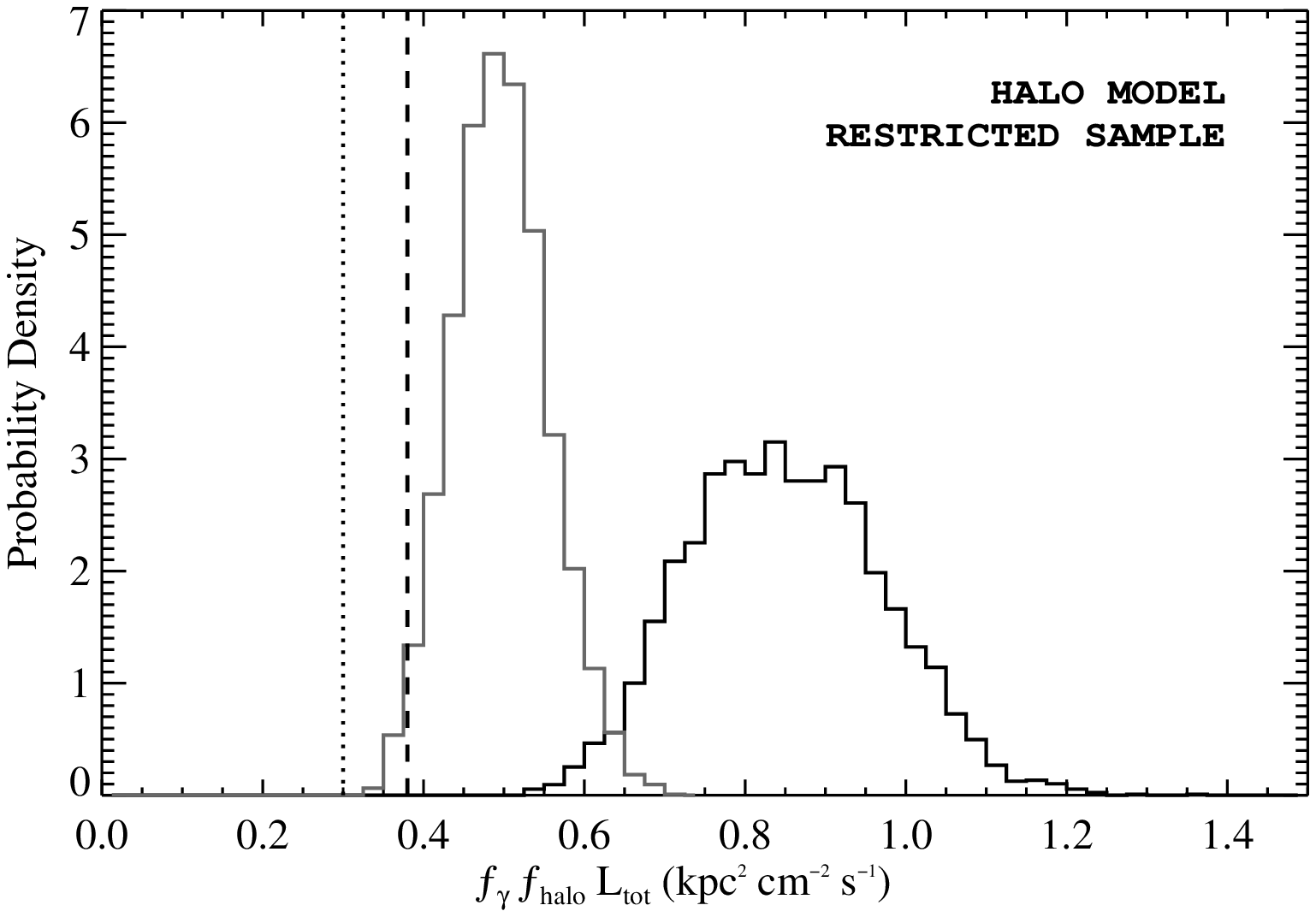}
\caption{Probability density function for total luminosity of a halo population of sources.  The resulting distribution for 5000 realizations using only sources with $|b|>5^\circ$ ({\it light gray}) is shown along with the corresponding distribution when all sources are used ({\it black}), drawing from the original sample ({\it left panel}) and the restricted sample ({\it right panel}).  The M31 limit on the total gamma-ray luminosity $\ltot$ ({\it dashed}) and derived limit for the luminosity of point sources $\lpt$ ({\it dotted}) are also shown.\label{fig:haloresults}}
\end{figure}

Figure \ref{fig:haloresults} shows distributions of the total source luminosity for a halo population.  Following a similar approach to that for the disk and bulge distributions, for this scenario we evaluate the total population luminosity for both of our source samples by first restricting each sample to sources with $|b|>5^\circ$ (omitting sources which appear to be in the Galactic plane), and then considering all sources in each sample.  Assuming a Gaussian distribution, the mean value for the original sample using sources with $|b|>5^\circ$ is $\mu = 0.64$, in units of (kpc/cm)$^{2}$ s$^{-1}$.  Without the latitude restriction, the distribution shifts up to $\mu = 1.7$.  Our more conservative, restricted sample also exceeds the M31 limits in almost all realizations, with $\mu = 0.49$ for sources with $|b|>5^\circ$, and $\mu = 0.84$ for all sources in this sample.  Based on these results, the possibility of any halo distribution of these sets of unidentified sources producing a luminosity below even the maximum observational limit is strongly disfavored.

\section{Discussion}
\label{fermi}

With the launch of {\it Fermi} last year, this is an ideal time to consider source identification strategies.  
The number of detected gamma-ray sources will be increased substantially by {\it Fermi}, and with it the number of sources to be identified with known low-energy counterparts.  With {\it Fermi}'s expanded source catalog, following each detection with multiwavelength observations will certainly be an impractical method for making identifications.  Although other gamma-ray telescopes currently in operation such as HESS, MAGIC, and VERITAS will be able to use their superior angular resolution and variability measurements to help identify sources \citep[e.g.,][]{reimer_funk_07}, they face similar issues with regard to the burden of multiwavelength follow-up observations of large numbers of objects not easily identified by other methods.  In this work we proposed an additional constraint to aid in identifying populations of sources which is complementary to constraints from existing approaches.

Of particular interest for this study is the likely detection of M31 by {\it Fermi} \citep{digel_etal_00,pavlidou_fields_01}.  A measurement of the M31 gamma-ray flux will significantly narrow the range of plausible point source luminosities and provide stronger constraints on the properties of new Galactic gamma-ray emitting populations.  Furthermore, {\it Fermi} may be able to provide detailed spatial and spectral information about the gamma-ray emission of the Milky Way and M31, which would test the assumption of similarity underlying this work.  A determination of the M31 flux would also test the validity of the predicted diffuse gamma-ray emission.

As an independent test of the plausibility of a candidate Galactic population, the method proposed here can be used in conjunction with isotropy studies to generate stronger constraints than can be achieved by isotropy arguments alone.  
Similarly, the luminosity constraint outlined here stands to benefit greatly from isotropy information, since satisfying the luminosity constraint alone does not guarantee that a proposed distribution is plausible.  Any realistic Galactic population is expected to be distributed according to symmetries of the Galaxy (e.g., it is unlikely that a population would cluster near our position).  Our method of assigning distances effectively projects all of the sources into the disk and bulge or into the halo, so additional constraints, such as those from isotropy studies, are necessary to  determine whether realizations of a population represent plausible spatial distributions.  We have shown here that most of the {EGRET} unidentified sources cannot be distributed in the halo if the M31 constraint is to be satisfied, so this is a particularly important issue for sources with high Galactic latitude which, naively, are least likely to be associated with the disk and bulge.  In general, as evident from Figures~\ref{fig:dbresults} and \ref{fig:haloresults}, this method is intrinsically weaker for sources associated with the disk and bulge, which are typically much nearer to our position than halo sources.

It is important to keep in mind that although we expect M31 and the Milky Way to be similar in gamma-rays, clearly we do not expect them to be identical.  Depending on the characteristics of the source population and the properties of M31, we expect that the total population luminosities we calculated by placing the unidentified sources in the Milky Way could vary by a factor of a few for a corresponding population in M31.  

There is currently only an upper limit to the M31 flux, so a flux measurement by {\it Fermi} would necessarily affirm or tighten the luminosity constraint.  Once a determination of the M31 flux is made, it is possible that no new Galactic populations could be accommodated, unless M31 and the Milky Way are fundamentally different in their gamma-ray emission properties.  Furthermore, a flux determination would also suggest a lower limit to the point source flux.  
In this study, we did not take into account a number of probable contributors to the Galactic luminosity: diffuse emission from unresolved point sources, identified Milky Way objects, and unidentified sources with recently suggested Galactic counterparts.  We expect that M31 will have similar populations of these Galactic sources, so in principle this will decrease the allowed luminosity of new proposed populations, and consequently further constrain their abundance and distribution.  

\section{Conclusions}
\label{conc}

We outlined a new approach to constraining the origin of unidentified gamma-ray sources based on a luminosity constraint.  This approach provides a constraint on candidate populations which is complementary to, and can be used in conjunction with, constraints from existing approaches.  
We demonstrated this method by applying it to the unidentified sources in the third {EGRET} catalog.  Using the observational upper limit for the gamma-ray luminosity of M31 along with the assumption that M31 and the Milky Way host similar gamma-ray emitting populations, we constrained the allowed spatial distribution of these sources. 
 
We found that it is highly unlikely that a substantial fraction of the {EGRET} unidentified sources are members of a Galactic halo population.  For the case of source populations expected to be associated with the disk and bulge, we find that within reasonable uncertainties all of the {EGRET} unidentified sources can be of Galactic origin without exceeding our assumed luminosity upper limit, although isotropy studies could likely further constrain this scenario.  

The new information {\it Fermi} brings, with respect to both known gamma-ray sources and possible detections of proposed emitters, will help to determine the nature of the unidentified sources.  Statistical techniques such as those employed in this study will be particularly useful in making meaningful statements about gamma-ray emitting populations when large numbers of individual identifications are not feasible.

\ack
We thank J. Beacom, A. Strong, O. Reimer, and I. Grenier for insightful comments and suggestions relating to this work, and D. Thompson for his advice with regard to the observational upper limits from the {EGRET} data.  We are extremely grateful to S. Digel for detailed discussions about the {EGRET} upper limits and diffuse backgrounds.  This work was supported by the Kavli Institute for Cosmological Physics at the University of Chicago through grants NSF PHY-0114422 and NSF PHY-0551142 and an endowment from the Kavli Foundation and its founder Fred Kavli.
Support for VP was provided by NASA through the GLAST Fellowship Program, NASA Cooperative Agreement: NNG06DO90A.

\bibliographystyle{unsrtnat}
\bibliography{bibliography}   

\begin{thebibliography}{55}
\providecommand{\natexlab}[1]{#1}
\providecommand{\url}[1]{\texttt{#1}}
\expandafter\ifx\csname urlstyle\endcsname\relax
  \providecommand{\doi}[1]{doi: #1}\else
  \providecommand{\doi}{doi: \begingroup \urlstyle{rm}\Url}\fi

\bibitem[{Hartman} et~al.(1999){Hartman}, {Bertsch}, {Bloom}, {Chen},
  {Deines-Jones}, {Esposito}, {Fichtel}, {Friedlander}, {Hunter}, {McDonald},
  {Sreekumar}, {Thompson}, {Jones}, {Lin}, {Michelson}, {Nolan}, {Tompkins},
  {Kanbach}, {Mayer-Hasselwander}, {M{\"u}cke}, {Pohl}, {Reimer}, {Kniffen},
  {Schneid}, {von Montigny}, {Mukherjee}, and {Dingus}]{hartman_etal_99}
R.~C. {Hartman}, D.~L. {Bertsch}, S.~D. {Bloom}, A.~W. {Chen},
  P.~{Deines-Jones}, J.~A. {Esposito}, C.~E. {Fichtel}, D.~P. {Friedlander},
  S.~D. {Hunter}, L.~M. {McDonald}, P.~{Sreekumar}, D.~J. {Thompson}, B.~B.
  {Jones}, Y.~C. {Lin}, P.~F. {Michelson}, P.~L. {Nolan}, W.~F. {Tompkins},
  G.~{Kanbach}, H.~A. {Mayer-Hasselwander}, A.~{M{\"u}cke}, M.~{Pohl},
  O.~{Reimer}, D.~A. {Kniffen}, E.~J. {Schneid}, C.~{von Montigny},
  R.~{Mukherjee}, and B.~L. {Dingus}.
\newblock {The Third EGRET Catalog of High-Energy Gamma-Ray Sources}.
\newblock \emph{ApJS}, 123:\penalty0 79--202, July 1999.
\newblock \doi{10.1086/313231}.

\bibitem[{Paredes} et~al.(2000){Paredes}, {Mart{\'{\i}}}, {Rib{\'o}}, and
  {Massi}]{paredes_etal_00}
J.~M. {Paredes}, J.~{Mart{\'{\i}}}, M.~{Rib{\'o}}, and M.~{Massi}.
\newblock {Discovery of a High-Energy Gamma-Ray-Emitting Persistent
  Microquasar}.
\newblock \emph{Science}, 288:\penalty0 2340--2342, June 2000.

\bibitem[{Berezinsky} et~al.(2003){Berezinsky}, {Dokuchaev}, and
  {Eroshenko}]{berezinsky_etal_03}
V.~{Berezinsky}, V.~{Dokuchaev}, and Y.~{Eroshenko}.
\newblock {Small-scale clumps in the galactic halo and dark matter
  annihilation}.
\newblock \emph{Phys.~Rev.~D}, 68\penalty0 (10):\penalty0 103003--+, November
  2003.

\bibitem[{Bergstr{\"o}m} et~al.(1999){Bergstr{\"o}m}, {Edsj{\"o}}, {Gondolo},
  and {Ullio}]{bergstrom_etal_99}
L.~{Bergstr{\"o}m}, J.~{Edsj{\"o}}, P.~{Gondolo}, and P.~{Ullio}.
\newblock {Clumpy neutralino dark matter}.
\newblock \emph{Phys.~Rev.~D}, 59\penalty0 (4):\penalty0 043506, February 1999.

\bibitem[{Blasi} et~al.(2003){Blasi}, {Olinto}, and
  {Tyler}]{blasi_olinto_tyler_03}
P.~{Blasi}, A.~V. {Olinto}, and C.~{Tyler}.
\newblock {Detecting WIMPs in the microwave sky}.
\newblock \emph{Astroparticle Physics}, 18:\penalty0 649--662, March 2003.

\bibitem[{Calc{\'a}neo-Rold{\'a}n} and {Moore}(2000)]{calcaneoroldan_moore_00}
C.~{Calc{\'a}neo-Rold{\'a}n} and B.~{Moore}.
\newblock {Surface brightness of dark matter: Unique signatures of neutralino
  annihilation in the galactic halo}.
\newblock \emph{Phys.~Rev.~D}, 62\penalty0 (12):\penalty0 123005--+, December
  2000.

\bibitem[{Tasitsiomi} and {Olinto}(2002)]{tasitsiomi_olinto_02}
A.~{Tasitsiomi} and A.~V. {Olinto}.
\newblock {Detectability of neutralino clumps via atmospheric Cherenkov
  telescopes}.
\newblock \emph{Phys.~Rev.~D}, 66\penalty0 (8):\penalty0 083006--+, October
  2002.

\bibitem[{Taylor} and {Silk}(2003)]{taylor_silk_03}
J.~E. {Taylor} and J.~{Silk}.
\newblock {The clumpiness of cold dark matter: implications for the
  annihilation signal}.
\newblock \emph{MNRAS}, 339:\penalty0 505--514, February 2003.

\bibitem[{Ullio} et~al.(2002){Ullio}, {Bergstr{\"o}m}, {Edsj{\"o}}, and
  {Lacey}]{ullio_etal_02}
P.~{Ullio}, L.~{Bergstr{\"o}m}, J.~{Edsj{\"o}}, and C.~{Lacey}.
\newblock {Cosmological dark matter annihilations into {$\gamma$} rays: A
  closer look}.
\newblock \emph{Phys.~Rev.~D}, 66\penalty0 (12):\penalty0 123502--+, December
  2002.

\bibitem[{Bertone} et~al.(2005){Bertone}, {Zentner}, and
  {Silk}]{bertone_etal_05}
G.~{Bertone}, A.~R. {Zentner}, and J.~{Silk}.
\newblock {New signature of dark matter annihilations: Gamma rays from
  intermediate-mass black holes}.
\newblock \emph{Phys.~Rev.~D}, 72\penalty0 (10):\penalty0 103517--+, November
  2005.

\bibitem[{Bloom} et~al.(1997){Bloom}, {Hartman}, {Teraesranta}, {Tornikoski},
  and {Valtaoja}]{bloom_97}
S.~D. {Bloom}, R.~C. {Hartman}, H.~{Teraesranta}, M.~{Tornikoski}, and
  E.~{Valtaoja}.
\newblock {Possible New Identifications for EGRET Sources}.
\newblock \emph{\apjl}, 488:\penalty0 L23+, October 1997.
\newblock \doi{10.1086/310911}.

\bibitem[{Zook} et~al.(1997){Zook}, {Giammona}, {Unwin}, {Wehrle},
  {Terasranta}, {Valtaoja}, {Kidger}, and {Gonzalez-Perez}]{zook_97}
Z.~C. {Zook}, W.~J. {Giammona}, S.~C. {Unwin}, A.~E. {Wehrle}, H.~{Terasranta},
  E.~{Valtaoja}, M.~R. {Kidger}, and J.~N. {Gonzalez-Perez}.
\newblock {Radio counterparts of unidentified EGRET gamma-ray sources.}
\newblock \emph{\aj}, 114:\penalty0 1121--+, September 1997.
\newblock \doi{10.1086/118542}.

\bibitem[{Gehrels} and {Michelson}(1999)]{gehrels_michelson_99}
N.~{Gehrels} and P.~{Michelson}.
\newblock {GLAST: the next-generation high energy gamma-ray astronomy mission}.
\newblock \emph{Astroparticle Physics}, 11:\penalty0 277--282, June 1999.

\bibitem[{Combi} et~al.(1999){Combi}, {Romero}, and {Benaglia}]{crb_99}
J.~A. {Combi}, G.~E. {Romero}, and P.~{Benaglia}.
\newblock {A Search for Radio Counterparts of Southern Unidentified EGRET
  Sources}.
\newblock \emph{\aj}, 118:\penalty0 659--665, August 1999.
\newblock \doi{10.1086/300960}.

\bibitem[{Tornikoski} et~al.(2002){Tornikoski}, {L{\"a}hteenm{\"a}ki},
  {Lainela}, and {Valtaoja}]{tornik_02}
M.~{Tornikoski}, A.~{L{\"a}hteenm{\"a}ki}, M.~{Lainela}, and E.~{Valtaoja}.
\newblock {Possible Identifications for Southern EGRET Sources}.
\newblock \emph{\apj}, 579:\penalty0 136--147, November 2002.
\newblock \doi{10.1086/342673}.

\bibitem[{Fegan} et~al.(2005){Fegan}, {Badran}, {Bond}, {Boyle}, {Bradbury},
  {Buckley}, {Carter-Lewis}, {Catanese}, {Celik}, {Cui}, {Daniel}, {D'Vali},
  {de la Calle Perez}, {Duke}, {Falcone}, {Fegan}, {Finley}, {Fortson},
  {Gaidos}, {Gammell}, {Gibbs}, {Gillanders}, {Grube}, {Hall}, {Hall}, {Hanna},
  {Hillas}, {Holder}, {Horan}, {Jarvis}, {Jordan}, {Kenny}, {Kertzman},
  {Kieda}, {Kildea}, {Knapp}, {Kosack}, {Krawczynski}, {Krennrich}, {Lang}, {Le
  Bohec}, {Lessard}, {Linton}, {Lloyd-Evans}, {Milovanovic}, {McEnery},
  {Moriarty}, {Mukherjee}, {Muller}, {Nagai}, {Nolan}, {Ong}, {Pallassini},
  {Petry}, {Power-Mooney}, {Quinn}, {Quinn}, {Ragan}, {Rebillot}, {Reynolds},
  {Rose}, {Schroedter}, {Sembroski}, {Swordy}, {Syson}, {Vassiliev}, {Wakely},
  {Walker}, {Weekes}, and {Zweerink}]{fegan_05}
S.~J. {Fegan}, H.~M. {Badran}, I.~H. {Bond}, P.~J. {Boyle}, S.~M. {Bradbury},
  J.~H. {Buckley}, D.~A. {Carter-Lewis}, M.~{Catanese}, O.~{Celik}, W.~{Cui},
  M.~{Daniel}, M.~{D'Vali}, I.~{de la Calle Perez}, C.~{Duke}, A.~{Falcone},
  D.~J. {Fegan}, J.~P. {Finley}, L.~F. {Fortson}, J.~A. {Gaidos}, S.~{Gammell},
  K.~{Gibbs}, G.~H. {Gillanders}, J.~{Grube}, J.~{Hall}, T.~A. {Hall},
  D.~{Hanna}, A.~M. {Hillas}, J.~{Holder}, D.~{Horan}, A.~{Jarvis},
  M.~{Jordan}, G.~E. {Kenny}, M.~{Kertzman}, D.~{Kieda}, J.~{Kildea},
  J.~{Knapp}, K.~{Kosack}, H.~{Krawczynski}, F.~{Krennrich}, M.~J. {Lang},
  S.~{Le Bohec}, R.~W. {Lessard}, E.~{Linton}, J.~{Lloyd-Evans},
  A.~{Milovanovic}, J.~{McEnery}, P.~{Moriarty}, R.~{Mukherjee}, D.~{Muller},
  T.~{Nagai}, S.~{Nolan}, R.~A. {Ong}, R.~{Pallassini}, D.~{Petry},
  B.~{Power-Mooney}, J.~{Quinn}, M.~{Quinn}, K.~{Ragan}, P.~{Rebillot}, P.~T.
  {Reynolds}, H.~J. {Rose}, M.~{Schroedter}, G.~H. {Sembroski}, S.~P. {Swordy},
  A.~{Syson}, V.~V. {Vassiliev}, S.~P. {Wakely}, G.~{Walker}, T.~C. {Weekes},
  and J.~{Zweerink}.
\newblock {A Survey of Unidentified EGRET Sources at Very High Energies}.
\newblock \emph{\apj}, 624:\penalty0 638--655, May 2005.
\newblock \doi{10.1086/429123}.

\bibitem[{Torres} et~al.(2005){Torres}, {Dame}, and {Romero}]{tdr_05}
D.~F. {Torres}, T.~M. {Dame}, and G.~E. {Romero}.
\newblock {Status of the Connection between Unidentified Egret Sources and
  Supernova Remnants: The Case of Cta 1}.
\newblock \emph{\apss}, 297:\penalty0 393--398, June 2005.
\newblock \doi{10.1007/s10509-005-7698-3}.

\bibitem[{La Palombara} et~al.(2006){La Palombara}, {Mignani},
  {Hatziminaoglou}, {Schirmer}, {Bignami}, and {Caraveo}]{lapalombara_etal_06}
N.~{La Palombara}, R.~P. {Mignani}, E.~{Hatziminaoglou}, M.~{Schirmer}, G.~F.
  {Bignami}, and P.~{Caraveo}.
\newblock {XMM-Newton and ESO observations of the two unidentified
  {$\gamma$}-ray sources 3EG J0616-3310 and 3EG J1249-8330}.
\newblock \emph{A\&A}, 458:\penalty0 245--257, October 2006.

\bibitem[{Paredes} et~al.(2008){Paredes}, {Mart{\'{\i}}}, {Ishwara-Chandra},
  {Torres}, {Romero}, {Combi}, {Bosch-Ramon}, {Mu{\~n}oz-Arjonilla}, and
  {S{\'a}nchez-Sutil}]{paredes_etal_08}
J.~M. {Paredes}, J.~{Mart{\'{\i}}}, C.~H. {Ishwara-Chandra}, D.~F. {Torres},
  G.~E. {Romero}, J.~A. {Combi}, V.~{Bosch-Ramon}, A.~J. {Mu{\~n}oz-Arjonilla},
  and J.~R. {S{\'a}nchez-Sutil}.
\newblock {Radio detections towards unidentified variable EGRET sources}.
\newblock \emph{\aap}, 482:\penalty0 247--253, April 2008.
\newblock \doi{10.1051/0004-6361:20078299}.

\bibitem[{Reimer} and {Funk}(2007)]{reimer_funk_07}
O.~{Reimer} and S.~{Funk}.
\newblock {Demystifying an unidentified EGRET source by VHE gamma-ray
  observations}.
\newblock \emph{\apss}, 309:\penalty0 203--207, June 2007.
\newblock \doi{10.1007/s10509-007-9461-4}.

\bibitem[{Merck} et~al.(1996){Merck}, {Bertsch}, {Dingus}, {Esposito},
  {Fichtel}, {Fierro}, {Hartman}, {Hunter}, {Kanbach}, {Kniffen}, {Lin},
  {Mayer-Hasselwander}, {Michelson}, {von Montigny}, {Muecke}, {Mukherjee},
  {Nolan}, {Pohl}, {Schneid}, {Sreekumar}, {Thompson}, and {Willis}]{merck_96}
M.~{Merck}, D.~L. {Bertsch}, B.~L. {Dingus}, J.~A. {Esposito}, C.~E. {Fichtel},
  J.~M. {Fierro}, R.~C. {Hartman}, S.~D. {Hunter}, G.~{Kanbach}, D.~A.
  {Kniffen}, Y.~C. {Lin}, H.~A. {Mayer-Hasselwander}, P.~F. {Michelson},
  C.~{von Montigny}, A.~{Muecke}, R.~{Mukherjee}, P.~L. {Nolan}, M.~{Pohl},
  E.~{Schneid}, P.~{Sreekumar}, D.~J. {Thompson}, and T.~D. {Willis}.
\newblock {Study of the spectral characteristics of unidentified galactic EGRET
  sources. Are they pulsar-like?}
\newblock \emph{\aaps}, 120:\penalty0 C465+, December 1996.

\bibitem[{Roberts} et~al.(2005){Roberts}, {Brogan}, {Gaensler}, {Hessels},
  {Ng}, and {Romani}]{roberts_etal_05}
M.~S.~E. {Roberts}, C.~L. {Brogan}, B.~M. {Gaensler}, J.~W.~T. {Hessels}, C.-Y.
  {Ng}, and R.~W. {Romani}.
\newblock {Pulsar Wind Nebulae in Egret Error Boxes}.
\newblock \emph{Ap\&SS}, 297:\penalty0 93--100, June 2005.

\bibitem[{Grenier}(2000)]{grenier_00}
I.~A. {Grenier}.
\newblock {EGRET Unidentified Sources}.
\newblock In B.~L. {Dingus}, M.~H. {Salamon}, and D.~B. {Kieda}, editors,
  \emph{American Institute of Physics Conference Series}, volume 515 of
  \emph{American Institute of Physics Conference Series}, pages 261--+, 2000.

\bibitem[{Reimer}(2001)]{reimerASSL01}
O.~{Reimer}.
\newblock {Gamma-Ray Properties of Unidentified EGRET Sources}.
\newblock In A.~{Carrami{\~n}ana}, O.~{Reimer}, and D.~J. {Thompson}, editors,
  \emph{The Nature of Unidentified Galactic High-Energy Gamma-Ray Sources},
  volume 267 of \emph{Astrophysics and Space Science Library}, pages 17--34,
  2001.

\bibitem[{Grenier}(2003)]{grenier03}
I.~A. {Grenier}.
\newblock {Unidentified EGRET sources in the Galaxy}.
\newblock In R.~{Bandiera}, R.~{Maiolino}, and F.~{Mannucci}, editors,
  \emph{Texas in Tuscany. XXI Symposium on Relativistic Astrophysics}, pages
  397--404, 2003.

\bibitem[{Zhang} et~al.(2004){Zhang}, {Collmar}, {Hermsen}, and
  {Sch{\"o}nfelder}]{zhang_etal_04}
S.~{Zhang}, W.~{Collmar}, W.~{Hermsen}, and V.~{Sch{\"o}nfelder}.
\newblock {Spectral constraints on unidentified EGRET gamma-ray sources from
  COMPTEL MeV observations}.
\newblock \emph{\aap}, 421:\penalty0 983--990, July 2004.
\newblock \doi{10.1051/0004-6361:20035671}.

\bibitem[{Nolan} et~al.(2003){Nolan}, {Tompkins}, {Grenier}, and
  {Michelson}]{nolan_etal_03}
P.~L. {Nolan}, W.~F. {Tompkins}, I.~A. {Grenier}, and P.~F. {Michelson}.
\newblock {Variability of EGRET Gamma-Ray Sources}.
\newblock \emph{ApJ}, 597:\penalty0 615--627, November 2003.

\bibitem[{Sturner} and {Dermer}(1995)]{stu_der_95}
S.~J. {Sturner} and C.~D. {Dermer}.
\newblock {Association of unidentified, low latitude EGRET sources with
  supernova remnants.}
\newblock \emph{\aap}, 293:\penalty0 L17--L20, January 1995.

\bibitem[{Colafrancesco}(2002)]{colaf_02}
S.~{Colafrancesco}.
\newblock {First gamma-rays from galaxy clusters. Preliminary evidence of the
  association of galaxy clusters with EGRET unidentified gamma-ray sources}.
\newblock \emph{\aap}, 396:\penalty0 31--51, December 2002.
\newblock \doi{10.1051/0004-6361:20020328}.

\bibitem[{Torres} and {Reimer}(2005)]{torres_reimer_05}
D.~F. {Torres} and O.~{Reimer}.
\newblock {A Systematic and Quantitative Approach to the Identification of
  High-Energy {$\gamma$}-Ray Source Populations}.
\newblock \emph{\apjl}, 629:\penalty0 L141--L144, August 2005.
\newblock \doi{10.1086/447765}.

\bibitem[{Reimer} and {Thompson}(2001)]{reimer_thompson_01}
O.~{Reimer} and D.~J. {Thompson}.
\newblock {LogN--LogS Studies of EGRET Sources}.
\newblock In \emph{International Cosmic Ray Conference}, volume~6 of
  \emph{International Cosmic Ray Conference}, pages 2566--+, August 2001.

\bibitem[{Grenier}(2001)]{grenier_01}
I.~A. {Grenier}.
\newblock {Stable EGRET Unidentified Sources in Regions of Star Formation}.
\newblock In F.~A. {Aharonian} and H.~J. {V{\"o}lk}, editors, \emph{American
  Institute of Physics Conference Series}, volume 558 of \emph{American
  Institute of Physics Conference Series}, pages 191--+, 2001.

\bibitem[{Bhattacharya} et~al.(2003){Bhattacharya}, {Aky{\"u}z}, {Miyagi},
  {Samimi}, and {Zych}]{bhatta_03}
D.~{Bhattacharya}, A.~{Aky{\"u}z}, T.~{Miyagi}, J.~{Samimi}, and A.~{Zych}.
\newblock {On the distribution of EGRET unidentified sources in the Galactic
  plane}.
\newblock \emph{\aap}, 404:\penalty0 163--170, June 2003.
\newblock \doi{10.1051/0004-6361:20030393}.

\bibitem[{Reimer}(2005)]{reimer_05}
O.~{Reimer}.
\newblock {On The Origin Of Unidentified EGRET Gamma-Ray Sources}.
\newblock In F.~A. {Aharonian}, H.~J. {V{\"o}lk}, and D.~{Horns}, editors,
  \emph{High Energy Gamma-Ray Astronomy}, volume 745 of \emph{American
  Institute of Physics Conference Series}, pages 184--198, February 2005.

\bibitem[{Pavlidou} et~al.(2008){Pavlidou}, {Siegal-Gaskins}, {Fields},
  {Olinto}, and {Brown}]{pavlidou_siegal-gaskins_fields_etal_08}
V.~{Pavlidou}, J.~M. {Siegal-Gaskins}, B.~D. {Fields}, A.~V. {Olinto}, and
  C.~{Brown}.
\newblock {Unresolved Unidentified Source Contribution to the Gamma-Ray
  Background}.
\newblock \emph{\apj}, 677:\penalty0 27--36, April 2008.
\newblock \doi{10.1086/524723}.

\bibitem[{Mukherjee} et~al.(1995){Mukherjee}, {Bertsch}, {Dingus}, {Kanbach},
  {Kniffen}, {Sreekumar}, and {Thompson}]{mukh_95}
R.~{Mukherjee}, D.~L. {Bertsch}, B.~L. {Dingus}, G.~{Kanbach}, D.~A. {Kniffen},
  P.~{Sreekumar}, and D.~J. {Thompson}.
\newblock {On the nature of the unidentified EGRET sources: Are they
  Geminga-like pulsars?}
\newblock \emph{\apjl}, 441:\penalty0 L61--L64, March 1995.
\newblock \doi{10.1086/187790}.

\bibitem[{Klypin} et~al.(2002){Klypin}, {Zhao}, and {Somerville}]{kzs_02}
A.~{Klypin}, H.~{Zhao}, and R.~S. {Somerville}.
\newblock {{$\Lambda$}CDM-based Models for the Milky Way and M31. I. Dynamical
  Models}.
\newblock \emph{ApJ}, 573:\penalty0 597--613, July 2002.

\bibitem[{Blom} et~al.(1999){Blom}, {Paglione}, and
  {Carrami{\~n}ana}]{blom_etal_99}
J.~J. {Blom}, T.~A.~D. {Paglione}, and A.~{Carrami{\~n}ana}.
\newblock {Diffuse Gamma-Ray Emission from Starburst Galaxies and M31}.
\newblock \emph{ApJ}, 516:\penalty0 744--749, May 1999.

\bibitem[{McConnachie} et~al.(2005){McConnachie}, {Irwin}, {Ferguson}, {Ibata},
  {Lewis}, and {Tanvir}]{mcconnachie_etal_05}
A.~W. {McConnachie}, M.~J. {Irwin}, A.~M.~N. {Ferguson}, R.~A. {Ibata}, G.~F.
  {Lewis}, and N.~{Tanvir}.
\newblock {Distances and metallicities for 17 Local Group galaxies}.
\newblock \emph{MNRAS}, 356:\penalty0 979--997, January 2005.
\newblock \doi{10.1111/j.1365-2966.2004.08514.x}.

\bibitem[{Pavlidou} and {Fields}(2001)]{pavlidou_fields_01}
V.~{Pavlidou} and B.~D. {Fields}.
\newblock {Diffuse Gamma Rays from Local Group Galaxies}.
\newblock \emph{ApJ}, 558:\penalty0 63--71, September 2001.

\bibitem[{Hunter} et~al.(1997){Hunter}, {Bertsch}, {Catelli}, {Dame}, {Digel},
  {Dingus}, {Esposito}, {Fichtel}, {Hartman}, {Kanbach}, {Kniffen}, {Lin},
  {Mayer-Hasselwander}, {Michelson}, {von Montigny}, {Mukherjee}, {Nolan},
  {Schneid}, {Sreekumar}, {Thaddeus}, and
  {Thompson}]{hunter_bertsch_catelli_etal_97}
S.~D. {Hunter}, D.~L. {Bertsch}, J.~R. {Catelli}, T.~M. {Dame}, S.~W. {Digel},
  B.~L. {Dingus}, J.~A. {Esposito}, C.~E. {Fichtel}, R.~C. {Hartman},
  G.~{Kanbach}, D.~A. {Kniffen}, Y.~C. {Lin}, H.~A. {Mayer-Hasselwander}, P.~F.
  {Michelson}, C.~{von Montigny}, R.~{Mukherjee}, P.~L. {Nolan}, E.~{Schneid},
  P.~{Sreekumar}, P.~{Thaddeus}, and D.~J. {Thompson}.
\newblock {EGRET Observations of the Diffuse Gamma-Ray Emission from the
  Galactic Plane}.
\newblock \emph{\apj}, 481:\penalty0 205--+, May 1997.
\newblock \doi{10.1086/304012}.

\bibitem[{Prada} et~al.(2006){Prada}, {Klypin}, {Simonneau}, {Betancort-Rijo},
  {Patiri}, {Gottl{\"o}ber}, and
  {Sanchez-Conde}]{prada_klypin_simonneau_etal_06}
F.~{Prada}, A.~A. {Klypin}, E.~{Simonneau}, J.~{Betancort-Rijo}, S.~{Patiri},
  S.~{Gottl{\"o}ber}, and M.~A. {Sanchez-Conde}.
\newblock {How Far Do They Go? The Outer Structure of Galactic Dark Matter
  Halos}.
\newblock \emph{\apj}, 645:\penalty0 1001--1011, July 2006.
\newblock \doi{10.1086/504456}.

\bibitem[{Diemand} et~al.(2008){Diemand}, {Kuhlen}, {Madau}, {Zemp}, {Moore},
  {Potter}, and {Stadel}]{diemand_kuhlen_madau_etal_08}
J.~{Diemand}, M.~{Kuhlen}, P.~{Madau}, M.~{Zemp}, B.~{Moore}, D.~{Potter}, and
  J.~{Stadel}.
\newblock {Clumps and streams in the local dark matter distribution}.
\newblock \emph{\nat}, 454:\penalty0 735--738, August 2008.
\newblock \doi{10.1038/nature07153}.

\bibitem[{Springel} et~al.(2008){Springel}, {White}, {Frenk}, {Navarro},
  {Jenkins}, {Vogelsberger}, {Wang}, {Ludlow}, and
  {Helmi}]{springel_white_frenk_etal_08}
V.~{Springel}, S.~D.~M. {White}, C.~S. {Frenk}, J.~F. {Navarro}, A.~{Jenkins},
  M.~{Vogelsberger}, J.~{Wang}, A.~{Ludlow}, and A.~{Helmi}.
\newblock {Prospects for detecting supersymmetric dark matter in the Galactic
  halo}.
\newblock \emph{\nat}, 456:\penalty0 73--76, November 2008.
\newblock \doi{10.1038/nature07411}.

\bibitem[{Sowards-Emmerd} et~al.(2003){Sowards-Emmerd}, {Romani}, and
  {Michelson}]{sowards-emmerd_romani_michelson_03}
D.~{Sowards-Emmerd}, R.~W. {Romani}, and P.~F. {Michelson}.
\newblock {The Gamma-Ray Blazar Content of the Northern Sky}.
\newblock \emph{\apj}, 590:\penalty0 109--122, June 2003.
\newblock \doi{10.1086/374981}.

\bibitem[{Sowards-Emmerd} et~al.(2004){Sowards-Emmerd}, {Romani}, {Michelson},
  and {Ulvestad}]{sowards-emmerd_romani_michelson_etal_04}
D.~{Sowards-Emmerd}, R.~W. {Romani}, P.~F. {Michelson}, and J.~S. {Ulvestad}.
\newblock {Blazar Counterparts for 3EG Sources at -40 deg < Decl. < 0 deg:
  Pushing South through the Bulge}.
\newblock \emph{\apj}, 609:\penalty0 564--575, July 2004.
\newblock \doi{10.1086/421239}.

\bibitem[{Casandjian} and {Grenier}(2008)]{casandjian_grenier_08}
J.-M. {Casandjian} and I.~A. {Grenier}.
\newblock {A revised catalogue of EGRET {$\gamma$}-ray sources}.
\newblock \emph{\aap}, 489:\penalty0 849--883, October 2008.
\newblock \doi{10.1051/0004-6361:200809685}.

\bibitem[{Abdo}(2009)]{abdo_09}
A.~A. {Abdo}.
\newblock {Fermi Large Area Telescope Bright Gamma-ray Source List}.
\newblock \emph{ArXiv e-prints}, February 2009.

\bibitem[{Miyamoto} and {Nagai}(1975)]{miyamoto_nagai_75}
M.~{Miyamoto} and R.~{Nagai}.
\newblock {Three-dimensional models for the distribution of mass in galaxies}.
\newblock \emph{PASJ}, 27:\penalty0 533--543, 1975.

\bibitem[{Hernquist}(1990)]{hernquist_90}
L.~{Hernquist}.
\newblock {An analytical model for spherical galaxies and bulges}.
\newblock \emph{ApJ}, 356:\penalty0 359--364, June 1990.

\bibitem[{Johnston} et~al.(1996){Johnston}, {Hernquist}, and
  {Bolte}]{johnston_etal_96}
K.~V. {Johnston}, L.~{Hernquist}, and M.~{Bolte}.
\newblock {Fossil Signatures of Ancient Accretion Events in the Halo}.
\newblock \emph{ApJ}, 465:\penalty0 278--+, July 1996.

\bibitem[{Navarro} et~al.(1995){Navarro}, {Frenk}, and {White}]{nfw_95}
J.~F. {Navarro}, C.~S. {Frenk}, and S.~D.~M. {White}.
\newblock {Simulations of X-ray clusters}.
\newblock \emph{MNRAS}, 275:\penalty0 720--740, August 1995.

\bibitem[{Mattox} et~al.(2001){Mattox}, {Hartman}, and
  {Reimer}]{mattox_etal_01}
J.~R. {Mattox}, R.~C. {Hartman}, and O.~{Reimer}.
\newblock {A Quantitative Evaluation of Potential Radio Identifications for 3EG
  EGRET Sources}.
\newblock \emph{ApJS}, 135:\penalty0 155--175, August 2001.

\bibitem[{Grenier}(1997)]{grenier_97}
I.~A. {Grenier}.
\newblock {On the Nature of the Nearby Unidentified Gamma-Ray Sources}.
\newblock In C.~{Winkler}, T.~J.-L. {Courvoisier}, and P.~{Durouchoux},
  editors, \emph{ESA SP-382: The Transparent Universe}, pages 187--+, 1997.

\bibitem[{Digel} et~al.(2000){Digel}, {Moskalenko}, {Ormes}, {Sreekumar}, and
  {Williamson}]{digel_etal_00}
S.~{Digel}, I.~V. {Moskalenko}, J.~F. {Ormes}, P.~{Sreekumar}, and
  R.~{Williamson}.
\newblock {What can GLAST say about the origin of cosmic rays in other
  galaxies?}
\newblock In R.~A. {Mewaldt}, J.~R. {Jokipii}, M.~A. {Lee}, E.~{M{\"o}bius},
  and T.~H. {Zurbuchen}, editors, \emph{AIP Conf. Proc. 528}, pages 449--452,
  2000.

\end{thebibliography}

\end{document}